# Smart Home Crawler

## Towards a framework for semi-automatic IoT sensor integration

Martin Strohbach, Luis Adan Saavedra, Pavel Smirnov, Stefaniia Legostaieva

AGT International
Darmstadt, Germany

*Abstract*— Sensor deployments in Smart Homes have long reached commercial relevance for applications such as home automation, home safety or energy consumption awareness and reduction. Nevertheless, due to the heterogeneity of sensor devices and gateways, data integration is still a costly and time-consuming process. In this paper we propose the Smart Home Crawler Framework that (1) provides a common semantic abstraction from the underlying sensor and gateway technologies, and (2) accelerates the integration of new devices by applying machine learning techniques for linking discovered devices to a semantic data model. We present a first prototype that was demonstrated at ICT 2018. The prototype was built as a domain-specific crawling component for IoTCrawler, a secure and privacy-preserving search engine for the Internet of Things.

*Index Terms*— Internet of Things, Search Engines, Smart Home, Data Integration, Machine Learning

## I. INTRODUCTION

As typical for many IoT domains, Smart Home environments are extremely heterogeneous. Occupants buy and deploy sensors and actuators from different vendors each using different communication protocols such as ZigBee, Z-Wave, Bluetooth, WiFi, and each having different capabilities such as sensing and actuating. These devices are typically integrated by so-called home gateways that offer a single platform for accessing all connected devices. While the list of supported devices sounds partially impressive – e.g. Vera claims to be compatible with up to 2000 devices – existing gateways only contribute to solving the device heterogeneity problem in a very limited and restricted way. Each of the gateway vendors or open source projects follows their own approach of defining APIs and exhibits device metadata in different formats and to a varying level of detail. In short, there are no widely adopted Smart Home APIs that go beyond voice control. Consequently, application developers are forced to integrate with vendor-specific APIs requiring spending considerable time, money, and effort.

In combination with privacy concerns of home occupants, this could be a major roadblock towards realising innovative, AI-enabled Smart Home applications in existing domains of home automation, energy saving, security, elderly care and beyond.

In this paper we address the challenge of accelerating the integration of IoT sensors. We call our approach *IoT Crawling* that is generally applicable to other IoT domains. We define IoT Crawling as discovering and understanding the deployment of heterogeneous IoT devices and their data relating to a domain of interest such as a Smart Home, a city or a Smart Factory. Here we describe a concrete instance of a Smart Home Crawler that exploits machine learning to map devices to a semantic model. The Smart Home Crawler is implemented as a domain-specific crawler component of the cross-domain IoTCrawler framework. IoTCrawler is a secure and privacy-preserving search engine for the Internet of Things [6][9].

We first relate our ideas to existing research in Section II. In Section III we describe two Smart Home services as motivation for potential benefits of using the Smart Home Crawler. Then we describe our general framework in Section IV and present a concrete prototype that has been demonstrated at ICT 2018 event in Vienna, Austria. In Section V we discuss our current implementation and conclude with a summary in Section VI.

## II. RELATED WORK

We are not aware of any comparable approaches that integrate IoT sensors in a semi-automatic way. This paper seeks to address the general problem of semantic interoperability [13] using Semantic Web approaches as demonstrated by many semantic IoT systems such as OpenIoT[1]. However, existing systems do not generally follow the idea of an IoT search engine and leave the semantic annotation process to integrators. In our approach we explore the potential for using machine learning to partially automate IoT sensor integration.

In a broader sense, our work is also related to data integration and transformation tools such as Wrangler [8]. However, these tools purely focus on data integration, and do not provide solutions for discovering, accessing and semantically describing IoT devices and their deployment.

Closer to our work are smart home systems that seek to provide a common semantic abstraction facilitating the integration of existing Smart home devices. For instance, DOG, an ontology-powered OSGi Domotic Gateway [3] exploits semantic web technologies for vendor-independent command execution across heterogeneous technologies. Similarly, Kotis and Katasonov describe the concept of Semantic Smart Gateway as a "unified command bus" for executing commands across heterogeneous Smart Home installations [11]. They propose to use an online ontology-learning mechanism for describing entities for semi-automatic annotations generation and an online ontology-alignment mechanism for calculating similarities between concepts [12]. In contrast, our work focuses on providing abstractions for sensing devices.

---

[1] http://www.openiot.eu, last visited 23/04/2019

Closest to our work are approaches that automate or accelerate the integration of IoT sensors and their data. The Plaster framework [10] is one such example that provides a comprehensive approach for integrating raw metadata from a new building into the Brick schema. However, their approach is targeted to the domain of buildings and is not directly applicable to Smart Homes.

## III. SMART HOME SERVICES

In this section we briefly highlight two Smart Home services that illustrate the potential benefits of our approach.

### A. Recognising User Behaviour

Recognising the behaviour and activities of home occupants is a key enabler for many applications in the home. For instance, lighting and heating can be adjusted to users' activity. Other examples include detecting anomalies in the way people behave. For instance, a Smart Home service can monitor elderly occupants and notify caregivers or relatives about unusual behaviour ensuring their well-being.

When building such applications, one of the main challenges is to select and integrate appropriate sensors and developing algorithms that reliably detect activities, infer behaviours and inform about relevant anomalies. Consequently, these systems are usually developed in a way that ties the algorithms to a specific combination of sensors and Smart Home gateways. Typically, this leads to large efforts and costs when additional sensors need to be used or existing ones replaced, e.g. when less expensive or technically better alternatives appear on the market.

The Smart Home Crawler approach has the potential to significantly reduce the integration effort as (1) it relies on a common semantic abstraction that decouples applications from specific technologies, and (2) it accelerates the integration of new IoT sensors by semi-automatically mapping new devices to a semantic data model.

To a limited degree, the mapping of various Smart Home deployments may even offer the potential to provide behaviour-based services reusing existing Smart Home deployments.

### B. Energy Awareness

Providing a detailed overview of energy consumption in a home has the potential to reduce energy consumption leading both to monetary savings and being more environmentally friendly. Within the GrowSmarter project[2] we have developed a full system including a dashboard that collects energy measurements from smart plugs and provides home occupants a real-time view on their device-level energy consumption.

In the dashboard, occupants can get a detailed real-time view for each device, have access to high resolution historical data and obtain energy usage metrics. The dashboard provides a detailed analysis of individual devices' energy consumption and cost, both based on time and automatically detected usages. In addition, it is possible to compare devices. Our system can also automatically detect device types connected to a smart plug minimising labelling effort.

Currently, the GrowSmarter system supports collecting energy measurements from Fibaro Wall Plugs[3] connected via the Homee Gateway[4] or Pikkerton plugs[5] integrated via our own gateway.

We implemented the application using our own data analytics platform that provides a generic data ingestion format and allows efficient stream processing. With the approaches defined in this paper we expect that it will be easier to further decouple the applications, the sensor, and gateway technologies and integrate new IoT data sources faster.

## IV. SMART HOME CRAWLER

In this section we describe the IoTCrawler architecture, the Smart Home Crawler Framework and a prototype that we demonstrated at ICT 2018. A video of our prototype is available online [15].

### A. IoTCrawler Architecture

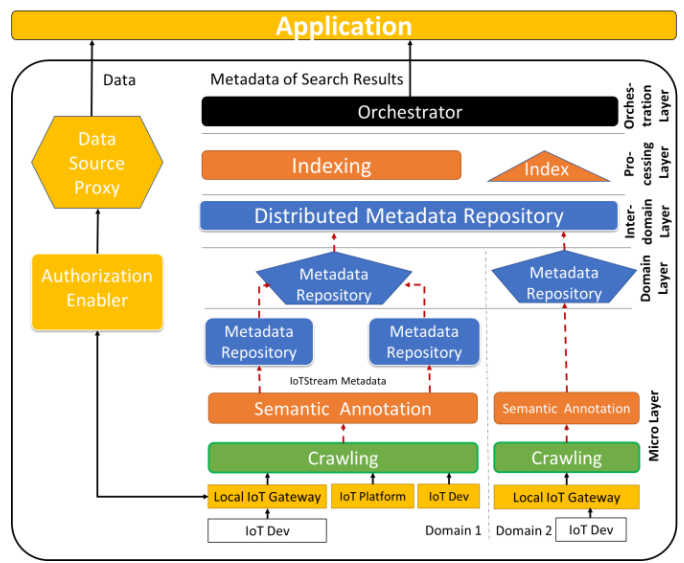

Fig. 1. Simplified view of the IoT Crawler Architecture

The Smart Home Crawler is a domain-specific component of the IoTCrawler search engine. The IoTCrawler infrastructure is designed in such a way that it focuses on managing metadata about connected IoT data sources rather than storing and managing the data itself. The metadata describes the IoT data source and is managed in a distributed, and hierarchically federated set of Metadata Repositories (MDR). The metadata is used to create an index for efficiently discovering data sources and for establishing a data path between the application and the data source (in our case the Smart Home gateway).

Fig. 1 shows a simplified view of the IoTCrawler infrastructure which is divided into several layers. The lowest, the micro layer, is closest to the data sources and contains the crawling function that discovers new data sources within a domain or subdomain. In a second step data sources are semantically annotated, and the resulting description is stored in the MDR. As can be seen from the figure, either an MDR in the micro-layer (e.g. deployed in the Smart home) or in the domain layer can be used. The MDR in the inter-domain provides a unified view of all available metadata in the repository. The MDR metadata is used to create an index based on which search queries issued by applications can be efficiently answered. In a first step, search queries return metadata in form of matching IoTStreams. In a second step, the application selects the corresponding sources and accesses the data source using an authorisation enabler that manages access to the data source and a data source proxy that exposes the data via a well-defined API.

---

[2] http://www.grow-smarter.eu/, last visited 11/04/2019
[3] https://www.fibaro.com/en/products/wall-plug/, last visited 11/04/2019
[4] https://hom.ee/, last visited 11/04/2019
[5] https://www.pikkerton.de/_objects/1/6.htm, last visited 11/04/2019

As depicted in Fig. 2, the metadata stored by the MDR is centred around the concepts of an IoTStream that extends the SSN ontology[6] to describe sensor streams.

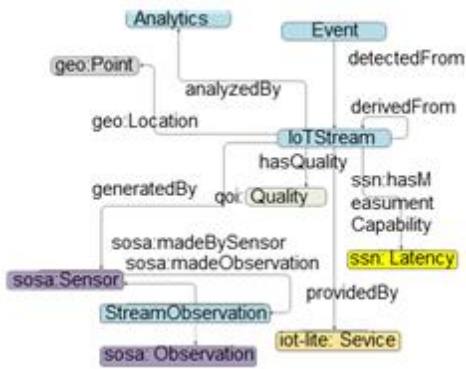

Fig. 2. The IoTCrawler IoTStream ontology

### B. Smart Home Crawler Framework

Fig. 3 shows a layered view of the Smart Home Crawler Framework using the same colour coding as in Fig. 1.

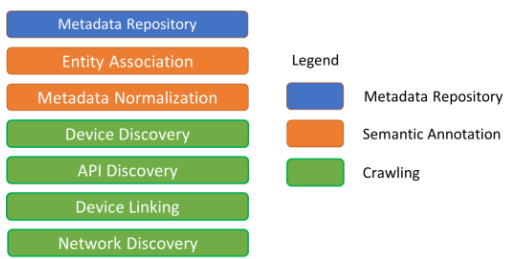

Fig. 3. Layers of the Smart Home Crawler Framework

The network discovery component is responsible for discovering local IoT devices that are connected to the local wired or wireless TCP/IP network and uses standard discovery protocols for discovering IoT devices such as home gateways and sensor devices connected to the home network.

The device linking component links observations from the network discovery agent to a device ontology that contains information about known devices such as their type and additional metadata such as vendor and product information.

The API discovery component is responsible for providing access to the metadata, i.e. metadata that describes how device data can be technically accessed (endpoint description, authorisation methods, etc.). A variety of methods can be used to obtain such information ranging from manually created descriptions to machine learning approaches to extract relevant API calls.

If the discovered device is a home gateway, the device discovery component uses the home gateway API to discover devices connected to the gateway. The device discovery concludes the actual crawling. Results are passed to the metadata normalisation component that normalises this metadata by linking it to an IoTStream representation. This process of metadata normalisation may also use sensor measurements.

Finally, the entity association component further enriches the metadata by linking devices to a domain level entity. Such a domain level entity may for instance be a vacuum cleaner that is connected to a smart plug that measures the energy consumed by the vacuum cleaner. In the SSN ontology such a domain level entity is called *Feature of Interest,* in the IoT reference model [1] this is called a *Virtual Entity*.

### C. Smart Home Crawler Prototype

Fig. 4 shows the architecture as implemented in our prototype. In the following sections we describe this architecture in detail using the component names of the layered structure of Fig. 3 and referencing the step in Fig. 4.

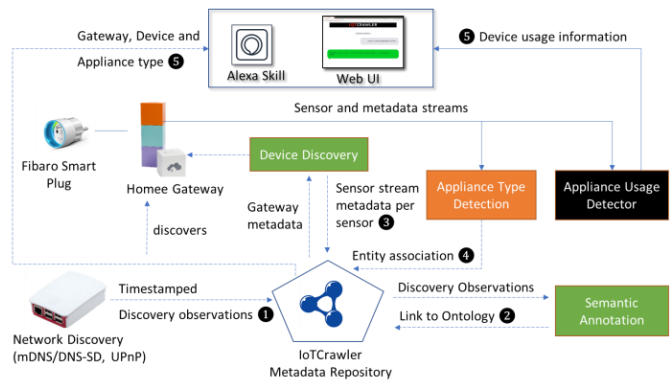

Fig. 4. Smart Home Crawler Architecture

*1) Metadata Repository:* The MDR is implemented using the AllegroGraph[7] triple stored deployed on an AWS EC2 instance.

*2) Network Discovery (Step 1):* our component scans local networks using UPnP, mDNS-SD and Bluetooth. The agent periodically scans the network and updates the triple store with timestamped discovery observations. These observations contain for instance information about the IP address, device names, and in the case of UPnP, also the manufacturer of the discovered devices and gateways. For instance, for a

---

[6] https://www.w3.org/TR/vocab-ssn/, last visited 11/04/2019

[7] https://franz.com/agraph/allegrograph/, last visited 11/04/2019

discovered Homee gateway we add the following triples[8]:
```
{
    "@id": "57bc95d6-4ed4-4b46-9101-f1d52871f872",
    "hasTimeStamp": "2018-10-29T12:13:01+01:00",
    "@label": "homee-0005510F1A3D",
    "hasNetworkName": "homee-0005510F1A3D",
}
```

*3) Device Linking (Step 2):* we query the triple store for discovered device names and try to discover a link to a class in our device ontology. Fig. 6 shows the relevant subset of our ontology.

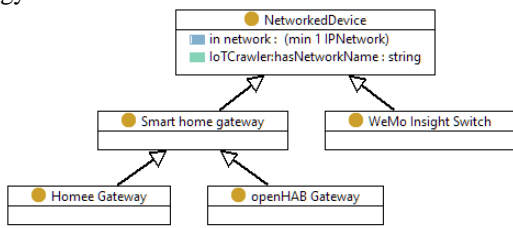

Fig. 5. Class hierarchy of relevant Device Ontology. Class names represent the values of the `rdfs:label` property that is used for linking.

We link the devices by performing a string similarity search between values of `hasNetworkName` and the values of properties in the device ontology. This way we can link the discovered network device to the `Homee Gateway` class by adding the triple:
```
{
  "@id" : "homee-0005510F1A3D",
  "@type" : "devices:HomeeGateway"
}
```
With that newlyadded triple we now have information about the homee Gateway as depicted in Fig. 6.

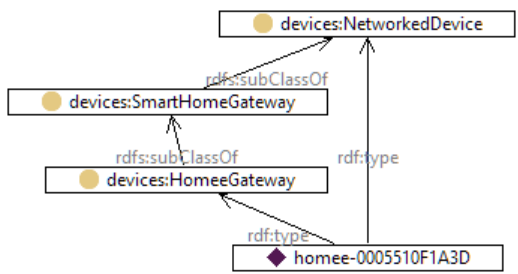

Fig. 6. As a result of the device linking step we know that the discovered device is a Homee Gateway (new `rdf:type` relationship between `homee-0005510F1A3D` and `HomeeGateway`.)

*4) API Discovery (not shown):* In order to obtain the endpoints that provide metadata and data about connected devices we have manually created adapters for Homee, OpenHAB[9] and Vera Secure[10]. However, we are also working on tooling for facilitating the identification of the relevant endpoints based on existing API documentations. For instance, we are able to extract a list of endpoints from API specifications such as Blueprint API and OpenAPI. A developer familiar with the interface can then semantically enrich the endpoints with information about which methods return metadata and data etc.

*5) Device Discovery (Step 3):* We use the gateway API to retrieve metadata about connected devices. The Homee gateway provides an API in which the metadata and measurements are accessible via WebSockets. The client obtains the device metadata by sending a `GET:nodes/` message. In the Homee API nodes represent devices connected to the gateway. Each node has a unique id and provides an exhaustive list of metadata items for the available nodes. Below we show a subset of the response sent by the gateway:

```
{"nodes":[{"added":1548863167,"id":7,"name":"FIBARO
System FGWPE/F Wall Plug Gen5"},{"added":1550568947,
"id":8,"name":"Fibaro%20Kitchen"}]}
```

Each node has a set of attributes that provides further data about the device. For instance, all information available for node 8 can be obtained by sending the message `GET:nodes/8`. Among other data the result consists of a list of attributes. Each attribute has a type. For the Fibaro plug, the homee gateway offers two attributes that refer to measurements: one of type `AccumulatedEnergyUse` and another one of type `CurrentEnergyUse`. The current energy use attribute is encoded as follows:

```
{"node":{"added":1550568947,"attributes":[{"current_
value":2.9,"id":64,"last_changed":1550570278,
"node_id":8,"type":3,"unit":"W"}]}}
```

As can be seen the key information about the semantics of the measurement is contained in the value of the `unit` name/value pair.

*6) Metadata Normalisation (Step 3):* In our current implementation we use the same approach as for device linking, i.e. we perform string similarity between the `name` value of the node object and the labels of the classes in the device ontology. This leads to the following triple being added to the triple store:
```
{
  "@id":
  "#58752baf-00f1-4fe2-a8df-8a097e5da983",
  "@type" : "devices:FibaroWallPlug",
  "label" : "Fibaro Kitchen"
}
```

Next, we compare the information from the `unit` value with units in the QUDT ontologies[11]. Using SSN ontology we can then express that the Fibaro Plug provides an `ElectricPowerObservation` that has a `Electric-PowerResult` which is a `qudt:QuantityValue` having `qudt:PowerUnit` as a unit.

In a similar way the Homee API provides "kWh" as a value for the `AccumulatedEnergyUse` attribute from which we can derive that the measurements relate to the physical quantity "energy" or "work".

---

[8] We use JSON-LD serialisation for showing RDF data. For brevity we omit the JSON-LD context.

[9] https://www.openhab.org/, last visited 11/04/2019

[10] https://getvera.com/, last visited 11/04/2019

[11] http://www.qudt.org/release2/qudt-catalog.html, last visited 11/04/2019

*7) Entity association (Step 4):* In our prototype, entity association means that we detect the type of an end user device or appliance connected to the smart plug. For that we reuse the Device Type Detection component used in the GrowSmarter project that currently can recognise 12 home appliances with a precision of 95%. Existing approaches [14] rely on multiple measurement types (e.g. active power, reactive power, current and/or voltage). However, in reality not all sensors can provide the required measurements. Therefore, we tested several algorithms that only require a stream of power measurements. The random forest classifier showed the best performance. The final result of device type detection is either a device type label or "not_confident".

*D. Activity Detection Prototype*

We have developed a simple activity detection application that detects user activities based on energy measurements of connected devices.

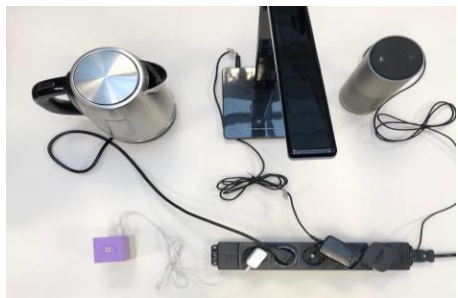

Fig. 7. Devices used in the demonstration. Bottom left: homee gateway, top (left to right): kettle and desk lamp connected to 2 Fibaro Wall Plug, Amazon Echo

As can be seen in Fig. 7, we use a desk lamp and a kettle, each connected to a "Fibaro Wall Plug" which communicates using the Z-Wave protocol with a Homee Gateway.

The application provides both a web-based and voice-controlled user interface implemented as an Alexa skill. For instance, the user can ask a question such as "*Alexa, ask Crawler what is happening at home*" based on which the system may reply "*It looks like someone is working at the table in the living room and somebody is boiling water in the kitchen*". The application can also be used for asking a range of questions related to network discovery, device discovery, appliance types (lamp, kettle) and device usage.

We infer activities based on usage detection of appliances connected to a Smart Plug, based on energy consumption thresholds. However, we plan to integrate and extend the more robust device usage detection as used in GrowSmarter.

## V. DISCUSSION

In the following subsections we shortly discuss some aspects of our current implementation.

*A. Implementation of the Metadata Repository*

As described above, we have used a triple store for the IoTCrawler Metadata Repository. This way we can reuse RDF vocabularies and ontologies and benefit from the powerful and standardised SPARQL query language. Despite ongoing research regarding scalability and federation of triple stores (e.g. [16]), there are still no mature and highly performant solutions. Therefore, we are currently examining the use of an NGSI-LD [7] broker that has been designed for distributed data management of IoT data [4][5]. It uses JSON-LD for exchanging data enabling semantic interoperability.

*B. Network Discovery*

In our implementation we have simply stored the results of the network device discovery results in our triple store. Modelling these results as IoT streams would allow exposing to applications. One such application could be a tool for crowd-based semantic annotations of the discovered data.

In this model the Network Discovery Agent would simply be modelled as a sensor that provides a stream of discovery observations. The data itself could be streamed using an NGSI-LD broker. For instance, we could use SSN for modelling the observations as follows:

```
{
  "@type": "DiscoveryObservation",
  "hasResult": {
    "@type": "DiscoveryResult",
    "discoveredDevice":"#homee-0005510F1A3D"
  }
  "resultTime": "2018-10-29T12:13:01+01:00"
}
```

*C. Device Discovery*

If an IoT device is connected to WLAN it may be discovered both during network discovery and device discovery using the gateway API. In this case we would try to determine whether the devices are the same. This can be reliably done if IP addresses are available in the metadata provided by the gateway API. If the devices cannot be distinguished, we simply create two instances. This means that as long as the ontology does not contain information from which it can be inferred that two devices instances are actually different (e.g. using `owl:inverseFunctional-Property`) the application should be prepared to have multiple representations of the same device.

*D. API Discovery*

In our implementation we rely on either explicitly defined semantic annotation of an API or well-structured API specifications such as OpenAPI or API Blueprint. In practice, however, well-structured API specifications are often not available. However, it may be possible to generate such specifications out of semi-structured online documentation [18], or to a limited degree using example API calls [17].

*E. Metadata Normalisation*

The described approach of using string similarity matching has limited applicability. For instance, in user-generated device names occupants typically encode locations (kitchen, living room) or associated appliances (toaster, lights) instead of the device model or its function. This further increases the complexity of metadata normalisation since user-generated data is highly language-dependent and often uses mixed languages. Consequently, we are considering making our detection approach language aware.

Another limitation is that data schemas often use abbreviations or encodings that require additional information for correctly interpreting their meaning as for instance required in the interpretation of the homee attribute types (see section IV.C.5). To solve this challenge, we are working on enriching our data with additional contextual information extracted from API documentation and vendor websites.

*F. Entity Association*

The simplest method to solve recognising device types is to ask users explicitly for the type of the attached appliance. However, based on our experience with various sensors and Smart Home systems deployed in various projects over several years, we have seen that (1) people do not necessarily provide required information, and (2) people provide information about entity associations implicitly, e.g. in device names, (3) even if there is correct metadata provided, Smart Home users change associations without necessarily updating the metadata. Consequently, it is important to use features from all three sources (explicit and structured, implicit and semi-structured, and ML on the sensor data) to maintain an updated picture over time.

In our current approach we have targeted the specific problem detecting the type of appliances connected to a metering smart plug. However, we believe that the general ML framework we are developing for this subset is applicable even beyond the Smart Home domain, e.g. for learning and automatically detecting the association between a wearable sensor and its wearer.

*G. Privacy, Trust, Security and User Empowerment*

While not addressed in this paper, it is crucial that home occupants have full control about their data and can decide what data is securely shared outside their homes, and which entities can access their data for what purpose. In the context of the IoTCrawler project, we are developing several mechanisms to address these issues. For instance, only data that needs to be accessible by IoTCrawler applications is shared using the MDR. In addition, partners in the project are applying CPABE [2] and smart contracts managed in blockchains in order to solve trust concerns between data providers and data consumers [6][9].

## VI. SUMMARY

In this paper we have outlined our initial work on Smart Home Crawling, an approach for faster integration of IoT sensor data in the Smart Home domain. We addressed the heterogeneity problem by building on the IoTCrawler architecture and proposing an initial framework for Smart Home Crawling that we believe can be generalised to other domains of IoT crawling and semi-automatic IoT data integration. Moreover, we have implemented an additional prototype and discussed future directions. We are now working on further developing and evaluating our methods for understanding IoT sensor data.

ACKNOWLEDGMENT

We thank all project partners, especially those who contributed to the IoTCrawler architecture and the IoTStream ontology. The work presented in this paper has partly been funded by the H2020 projects IoTCrawler and GrowSmarter under the grant agreement numbers 646456 and 779852.